\documentclass[twocolumn]{aastex631}
\usepackage{amsfonts,amsmath,graphicx,url,hyperref,nicefrac}
\usepackage{amssymb, verbatim, subfigure, paralist, soul, comment, multirow}
\usepackage{natbib}

\usepackage{color}

\shorttitle{Counter Jet Emission and X-ray Re-brightening 170817A} 
\shortauthors{Dastidar \& Duffell}

\begin{document}

\title{Could the recent rebrightening of the GW170817A afterglow be caused by a counter jet?}

\author[0009-0000-6548-6177]{Ranadeep G. Dastidar}
\affiliation{Department of Physics and Astronomy, Purdue University, 525 Northwestern Avenue, West Lafayette, IN 47907, USA}

\author[0000-0001-7626-9629]{Paul C. Duffell}
\affiliation{Department of Physics and Astronomy, Purdue University, 525 Northwestern Avenue, West Lafayette, IN 47907, USA}

\begin{abstract}

GRB170817A (also GW170817) became the first binary neutron star (BNS) merger event detected via gravitational waves and electromagnetic signals. Over the next 4 years, various multiband observations have led to re-imagine the various short Gamma Ray Burts (sGRB) and interstellar medium interaction models. While these models successfully explain the observed afterglow until $\sim$ 900 days, a re-brightening or excess flux was observed in the 1keV X-Ray band after $\sim$ 1000 days. In this study, we re-evaluate the jet parameters using the new observations (until $\sim$ 1234 days) with a boosted fireball jet model. We study the observable effects of the counter-jet for GRB170817A, using our new afterglow code, Firefly. Our results show that it is indeed possible for the observed excess to coincide with the emissions from the counter-jet ($\sim 800$ days). We also computed an empirical scaling law between the jet and counter-jet peak emission timescales and the observer angle. The Firefly code can also track the simulated object through the observers' sky and numerically model the apparent motion. The calculated apparent motion ($\approx 2.6 c$) does not match the observed apparent motion (7.5c to 5.2c). Hence we conclude, the excess flux of GRB170817A may not be associated with the counter jet; however, it is not enough to reject this hypothesis from the traditional counter jet visibility time scales, which predicts $\geq 5000$ days. The apparent motion, combined with the multi-band lightcurves, is needed to break degeneracy between geometrical parameters and the microphysical parameters of the afterglow.

\end{abstract}


\section{Introduction} \label{sec:intro}

Binary neutron star (BNS) mergers have long been believed to be a strong candidate for multimessenger astrophysics. On 2017 August 17, GW170817 (and GRB170817A) became the first such event to be detected directly. It provided unprecedented insights into the physical properties of pre-merger (GW) and post-merger (EM) BNS. The event was quickly localized \citep{Coulter+2017} to a nearby galaxy at 40.7 Mpc \citep{Cantiello+2018}. The initial EM spectrum ($\sim$ days) was powered by thermal emission from the merger ejecta (also known as the kilonova) and the nonthermal synchrotron emission dominating in the X-ray and radio bands.

The early kilonova emission, powered largely by radioactive decay of heavy chemical elements, was in agreement with theoretical predictions \citep{Metzger+2017}. Meanwhile, the nonthermal emission in the first $\sim$900 days has been associated with synchrotron emission from a GRB afterglow due to an ultrarelativistic jet pointing away from us \citep{Mooley+2018, Ghirlanda+2019, Hotokezaka+2019, Nathanail+2021}. The initial observations during this period did not show any spectral evolution across nine orders of magnitude of frequency, and they were characterized as synchrotron emission with power-law spectrum $F_{\nu} \propto \nu^{-(p-1)/2}$, with $p = 2.166 \pm 0.026$ \citep{Fong+2019, Hajela+2019, Troja+2020}. More recent observations, after 900 days since the merger, have found evidence for an excess in X-ray emission. The excess X-ray emission was measured with $L_X \approx 5\times 10^{38} \mbox{ erg s}^{-1}$ at 1234 days. However, similar observations at 3GHz (radio) and 5keV (X-ray) lacked such a strong excess \citep{Hajela+2022, Troja+2021}. 

The GRB afterglow has been studied extensively, and the physical parameters are constrained by several lightcurve and spectral fits \citep{Zhang+2009, Piran+2005} for $\delta t < 900$ days. Several simple and complex models of jets have also been employed to explain the behavior of synchrotron emission from GW170817A. Top-hat jet models, formed by angular truncation of a spherical Blandford-McKee blast wave solution \citep{Blandford+1976}, cannot account for the mild and steady rise of the nonthermal emission (\cite{Troja+2017}, and references therein). More complex models include, a jet-less fireball model \citep{Salafia+2017, Wu+2018}, the choked jet-cocoon model \citep{Nakar+2018}, and the choked jet-cocoon model with a fast tail \citep{Hotokezaka+2018} characterized by a mildly relativistic quasi-spherical outflow. Other models include a structured jet with wide-angle wings viewed off-axis (e.g. \cite{Kathirgamaraju+2017}, \cite{Margutti+2018} and other), like the Gaussian shaped jet model (\cite{Troja+2018} and references therein), the successful jet-cocoon model \citep{Duffell+2018}, and the boosted fireball model \citep{Wu+2018}. These complex models have been successful in explaining the spectral and temporal evolution of the synchrotron emission from GW170817A, but in the context of successful jet breakout, they are indistinguishable from each other (\cite{Wu+2018} and references therein). However, the lightcurves show significant statistical deviation from the recent X-ray excess observed at 1keV. This excess emission has been associated with kilonova ejecta \citep{Kathirgamaraju+2019}, and also compact object remnants like hypermassive NS, or a prompt collapse to a BH, or a spinning-down NS (\cite{Hajela+2022} and references therein).

In this paper, we adopt the boosted fireball model \citep{Duffell+2013}. We then revisit the parameter space for the standard external shock and ISM interaction model of GRB afterglows \citep{Sari+1999, Granot+2002} and include the most recent observations (up to \textless 1300 days). One should note that the existing best fit parameters are based on observations for $\delta t < 900$ days, and the lightcurves deviate beyond that epoch. We compare the re-evaluated jet and afterglow microphysical parameters with the existing best fits and the MCMC results. Motivated by the association of late-time bumps in afterglow lightcurves to counter jet re-brightening \citep{Granot+2003,Li+2004, Wang+2009, Zhang+2009}, we revisit this scenario as a probable source of the excess flux in the recent observations of GRB170817A. We further make predictions for later behaviour. We also explore the correlation between the ratio of the global peak flux to the counter jet bump and the observer angle.

The question we ask here is, being agnostic of the observer angle, is it possible to correlate the re-brightening timescale with the emergence of counter jet emissions? If so, what are the required conditions for it? The paper is presented as follows. In section \ref{sec:TheoMod}, we discuss the theoretical aspects behind the problem. This is followed by section \ref{sec:Numerical}, which describes the numerical implementations. Here we also introduce our new synchrotron radiation code \textit{Firefly}. This includes how we use \textit{Firefly} to map the flux distribution along the plane perpendicular to the observer's line of sight. We discuss our findings and overall line of thought in section \ref{sec:Results}. Section \ref{subsec:Countetjettime} discusses the empirical scaling law between the time scales of the forward jet peak emission and the counter jet peak emission. This is followed by a discussion comparing our simulated lightcurves to the observations in section \ref{sec:Lightcurve} and the spectrum in section \ref{sec:Spectrum}. Finally, we estimate the apparent superluminal motion for the brightest region of the GRB170817A afterglow through the sky in section \ref{sec:AppMotion}. We summarize and discuss our findings and their implication in the conclusions (section \ref{sec:conclusion}).

\section{Theoretical Models} \label{sec:TheoMod}

In this work, we use the boosted fireball jet model. The boosted fireball is a two-parameter model \citep{Duffell+2013} that generates a family of outflows after they have expanded many orders of magnitude larger than the merger scale. The two input parameters are $\eta_0 \sim E/M$ which is the fluid frame Lorentz factor of a blast with energy E and mass M, and $\gamma_B \sim \frac{1}{\theta_0}$ being the boost (in lab frame or blast frame) given to the said blast, and $\theta_0$ is the jet opening angle. In contrast to a conventional fireball which expands isotropically (with the Lorentz factor $\eta$ ,as per our convention), the boosted fireball has gets an external kick in a particular direction with Lorentz factor $\Gamma_B$. In the extreme limit $\Gamma_B \rightarrow 1$, the boosted fireball is the same as an conventional fireball. While on the other end for $\Gamma_B \rightarrow \infty$ it corresponds to an ultra-relativistic jet with a negligible jet opening angle ($\theta_0 \approx 1/\Gamma_B$). The explosion energy per fireball is $E_0 \sim \gamma_B\eta_0E$. Thus for a double sided jet, the total energy will be 
$2E_0$, which is related to the isotropic equivalent energy ($E_{iso}$) by the relation

\begin{equation}
\label{eq:Eiso}
    2E_0 \simeq 4\pi(\theta_0)^2\frac{dE}{d\Omega} \sim \frac{E_{iso}}{\gamma_B^2}
\end{equation}

A single fireball has previously been used to simulate the forward jet, and its dynamics are comparable to other standard jet models \citep{Wu+2018}. In this work we consider two symmetrically reflected fireballs along the jet axis, but they are boosted in the opposite directions. This simulates the joint evolution of the forward jet and the counter jet. We assume a constant and equal boost in both the directions. This is justified because the merger's spatial and temporal scales are much smaller than the afterglow's spatial and temporal scales.

For this study, we have used the standard GRB afterglow theory \citep{Sari+1998, Granot+2002}. This refers to models based on synchrotron radiation from a decelerating relativistic blast wave interacting with the interstellar medium. It assumes the radiation is generated by non-thermally distributed electrons accelerated by the forward shock. These electrons are further assumed to be distributed as a power-law of their Lorentz factor $n'_e = C_efn'\gamma_e'^{-p}, \mbox{ where } \gamma'_e>\gamma'_m$ \citep{Eerten+2013}. Where primed quantities are expressed in the fluid comoving frame, and \textit{f} is the fraction of electrons radiating ($n’$). Assuming electrons are accelerated to $\gamma’_e \rightarrow \infty$ this equation can be integrated to obtain the normalization constant $C_e$ as, $C_e = \frac{1}{f}(p-1)\gamma_m^{p-1}$. We found a marginal change in \textit{f} has no significant effect on the lightcurves. It was hence fixed at 1 for the rest of the analysis. The minimum electron Lorentz factor at which the accelerated electrons are radiating ($\gamma'_m$) is given by Eq. \ref{eq:GammaMin}. Similarly, the cooling break can be solved as in Eq. \ref{eq:gamc}, where $\epsilon_e$ and $\epsilon_B$ are the fraction of total energy ($\epsilon_{\rm{th}}$) converted to kinetic and magnetic energy respectively. Further, due to the relativistic nature of the blast wave, most of the density is concentrated in a very thin shell behind the forward shock. The radiation is dominated by electrons present in this shocked shell of width $\Delta R/R \sim 1/(12\Gamma^2)$, where $\Gamma$ is the jet head Lorentz factor.

\begin{equation}
\label{eq:epsB}
    \epsilon_B = \frac{B'^2}{8\pi e'_{th}}
\end{equation}

\begin{equation}
\label{eq:epsE}
    \epsilon_e = \frac{\int n'_e\gamma'_em_ec^2 d\gamma'_e}{e'_{th}}
\end{equation}

The quantities upstream (ahead of the shock, in the ISM) and downstream of the shock are related as, $n’_e = 4\Gamma n_e$ and $e’_{\rm{th}} = (\Gamma-1)n’_em_pc^2$. Thus the above equation after integrating and some algebra, can be solved to give,

\begin{equation}
\label{eq:epsB}
    B' = \sqrt{32\pi\Gamma(\Gamma-1)n_em_p\epsilon_Bc^2}
\end{equation}
and,
\begin{equation}
\label{eq:GammaMin}
    \gamma_m' = \frac{p-2}{p-1}\frac{\epsilon_em_p}{fm_e}(\Gamma-1)
\end{equation}

The corresponding break frequencies can be found using Eq. \ref{eq:nui}.

\begin{equation}
\label{eq:nui}
    \nu_i' = \frac{3eB'\gamma'^2_e}{4\pi m_e c}
\end{equation}

Where $\gamma'_i$ is the Lorentz factor in the local fluid frame, and \textit{i} corresponds to either cooling break or minimum Lorentz factor cut off. However beyond a certain critical Lorentz factor ($\gamma'_c$), the electrons starts loosing energy by cooling over a timescale $t'$ (in lab frame $t=\gamma t'$). The frequency at which this happens is hence called the cooling frequency, $\nu_c'$. This can be estimated by equating the power lost over some expansion time ($t’$) to the rest mass energy, and it comes out to be (see \cite{Eerten+2010} for detailed calculations):

\begin{equation}
\label{eq:gamc}
    \gamma'_c = \frac{6\pi m_e\gamma c}{\sigma_T B'^2t}
\end{equation}

The characteristic frequencies, $\nu_m$ and $\nu_c$, can be observed as two break frequencies in the spectrum. Thus in the local fluid frame, a given fluid element has monochromatic emissivity per unit volume:

\begin{equation}
\label{eq:epsE}
    \epsilon'_{\nu'} = \epsilon'_p\times
      \begin{cases}
      (\nu'/\nu'_m)^{1/3} & \text{if $\nu' < \nu'_m < \nu'_c$}\\
      (\nu'/\nu'_m)^{-(p-1)/2} & \text{if $\nu'_m < \nu' < \nu'_c$}\\
      (\nu'_c/\nu'_m)^{-(p-1)/2}(\nu'/\nu'_c)^{-p/2} & \text{if $\nu'_m < \nu'_c < \nu'$}\\
      (\nu'/\nu'_c)^{1/3} & \text{if $\nu' < \nu'_c < \nu'_m$}\\
      (\nu'/\nu'_c)^{-1/2} & \text{if $\nu'_c < \nu' < \nu'_m$}\\
      (\nu'_m/\nu'_c)^{-1/2}(\nu'/\nu'_m)^{-p/2} & \text{if $\nu'_c < \nu'_m < \nu'$}\\
      \end{cases}
\end{equation}

where,

\begin{equation}
    \epsilon'_p = \frac{\sqrt{3}e^3B'fn'}{m_ec^2}
\end{equation}

Since most of the mass for a relativistic blast wave is concentrated in a very thin shell behind the shock, the observed emission comes only from this optically thin shell. This has a volume element given by:

\begin{equation}
    dV = R^2\sin\theta dRd\theta d\phi
\end{equation}

This converts to observed flux from the lab frame for a fluid element having doppler factor $\delta$ as:

\begin{equation}
    F_{\nu}(\nu,t_{\rm{obs}}) = \frac{1+z}{4\pi D_L^2}\int\epsilon'_{\nu'}\times\delta^2\times dV
\end{equation}

Where \textit{z} is the cosmological redshift, and $D_L$ is luminosity distance of the source. And the local fluid frame frequency ($\nu'$) and time (\textit{t}$'$) are related to the observer frequency ($\nu_{\rm{obs}}$) and observer time ($t_{\rm{obs}}$) as:

\begin{equation}
    \nu_{\rm{obs}} = \frac{\delta}{1+z}\nu'
\end{equation}

\begin{equation}
    \frac{t_{\rm{obs}}}{(1+z)} = t'-\frac{\Vec{r}'\cdot\Hat{n}}{c}
\end{equation}

At any given time and frequency, the observed flux is a result of the total light collected from all the photons arriving at the same time to the observer. In the lab frame (or the center of blast frame), these contours are known as ``Equal Arrival Time Surfaces'' and are spread over the entire jet at all times. The two frames, lab frame and observer frame, can be bridged by taking a projection of the local fluid element ($\Vec{r}$) (with respect to the center of blast), along the line of sight of the observer ($\hat{n}$).

\begin{equation}
\label{eq:Doppler}
    \delta  = \Gamma^{-1}(1-\Vec{v}\cdot\Hat{n})^{-1}
\end{equation}

\section{Numerical Implementation} \label{sec:Numerical}

The entire process of generating lightcurves and spectra from jets can be divided into two parts. First we hydrodynamically evolve two oppositely directed relativistic jets through a constant density interstellar medium (ISM) originating from the center of our domain. The domain is assumed to be in the BNS merger rest frame, and the fireballs are boosted with respect to this frame by Lorentzian transformations (see \cite{Duffell+2013} for the detailed calculations). This was carried out using the two-dimensional relativistic hydrodynamics moving mesh code, \textit{JET} \citep{Duffell+2018}.

\begin{figure}
\centering
\includegraphics[width=.4\textwidth]{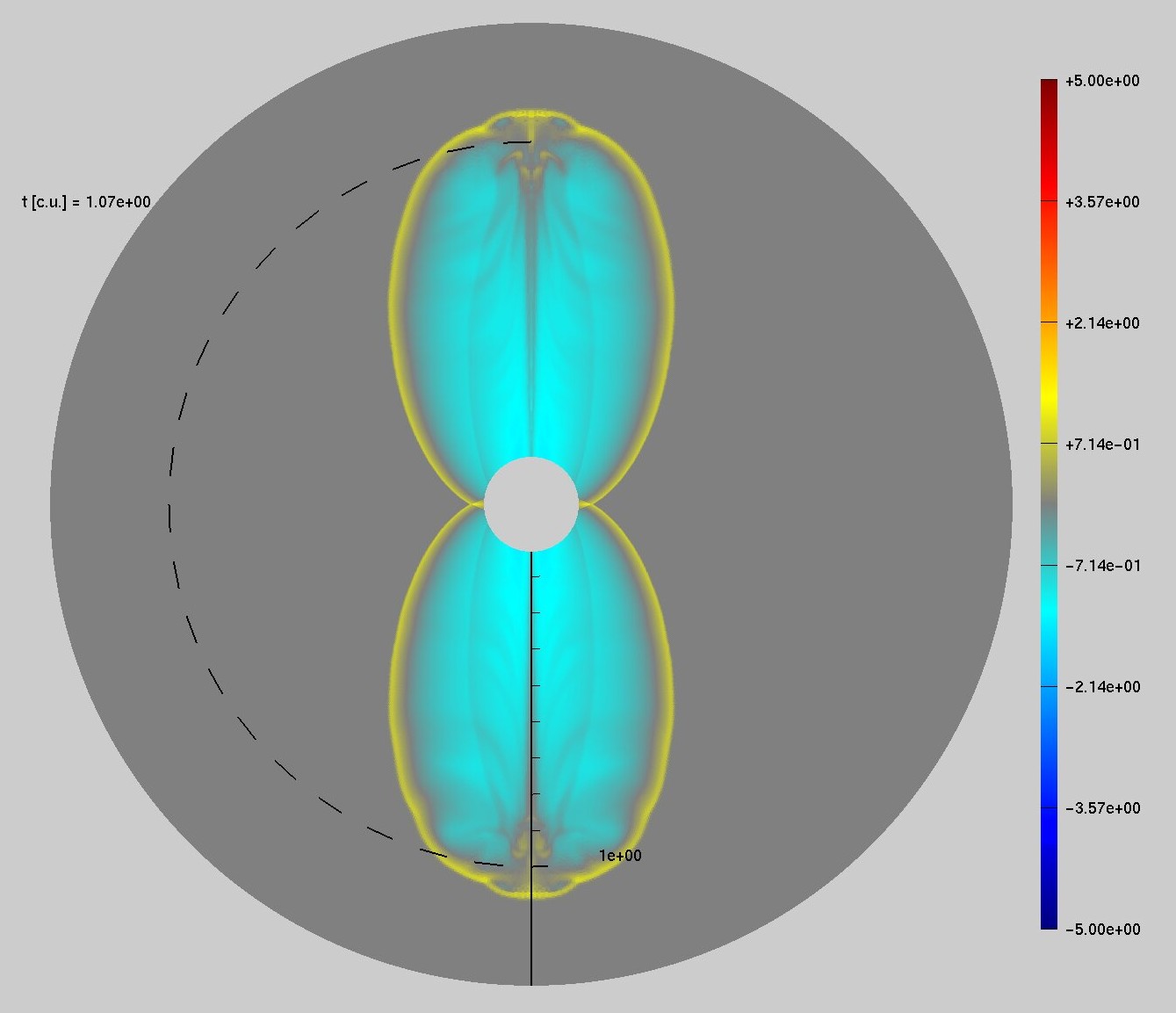}          
\caption{\footnotesize{Density snapshot of our jet-counterjet model using the JET code.  Two opposing boosted fireballs are injected, each with a single impulse normalized energy.  The jets are evolving through a constant density medium, $n_{\rm{\small{ISM}}} = 0.1$ cm$^{-3}$, marked by the grey background. The dashed black circle is for scaling reference, it marks the radial distance $r = 1.07 \mbox{[c.u.]} \approx 2.53\times 10^{18}$ cm.} The snapshot is the density profile at a typical time during its evolution.}
\label{fig:hydro}
\end{figure}

Each jet is initiated as a boosted fireball with a given fluid frame Lorentz factor $\eta_0$ and boosted with a Lorentz factor $\gamma_B$. The hydrodynamic simulations are carried out in code units normalized by setting the Sedov radius, $l \equiv (E/\rho_0)^{1/3}$, to unity, in the center of blast frame. This translates to the total blast energy E and a constant ISM density ($\rho_0$) set to unity. These are scaled to physical units during the afterglow calculations. The input for $\eta_0 \sim E/M$ sets the total ejected mass in the jet, and $\gamma_B \sim 1/\theta_0$ sets the jet opening angle ($\theta_0$), both in code units. The simulation begins around the time the fireball enters BM self similar phase, $t_{\rm{BM}} \sim (E/(\rho_0 c^5\Gamma^2))^{1/3}$, this sets the initial time $t_{\rm{min}} = 0.06t_{\rm{BM}}$ (where $c = 1$ in code units). This ensures that the blast wave evolution begins before the radiation is dominated by the ejecta swept up by the forward shock. The system is evolved until it expands to 20 times its Sedov radius (that is, $t_{\rm{final}} = 20l/c$), where it becomes subrelativistic. The system stratifies to a density profile given by Eq. \ref{eq:rho_ini}. A homologus expansion of the ejecta ($v = r/t$) is assumed.

 \begin{equation}\label{eq:rho_ini}
     \rho(r) = \rho_{\rm{max}}\left(\frac{1-R_{\rm{sh}}/t}{1-r/t}\right)
 \end{equation}

\begin{eqnarray}\label{eq:vel_ini}
      v(r,t) = 
  \begin{cases}
      r/t & \text{if $v < R_{\rm{sh}}$}\\
      0 & \text{otherwise}\\
      
  \end{cases}
\end{eqnarray}

\begin{equation}
    P \ll \rho
\end{equation}

 Where $\rho_{\rm{max}}\sim E/4\pi c^3t^3$ is the maximum blast wave density at the shock and follows from $E=4\pi\rho_{\rm{max}}(ct)^3c^2$. The shock radius ($R_{\rm{sh}}$) is given by $R_{\rm{sh}} = t(1-1/2\eta_0^2)$. The pressure is set to be very low ($10^{-5}\rho$) initially. An adiabatic equation of state is used with a constant adiabatic index of 4/3 for relativistic fluids. Although, by the end of our simulation the flow becomes subrelativistic, the afterglow emissions are dominant in the relativistic phase. The relativistic adiabatic constant holds true for marginally relativistic flow as well. Hence, a single value for the adiabatic constant does not affect the results significantly. The counter jet is implemented by reflecting the forward jet about the plane perpendicular to the jet axis. Lastly, the fireball is expanded in an interstellar medium of constant density normalized to unity.  A snapshot of the evolution is shown in Fig \ref{fig:hydro}.

Since the problem is axisymmetric, the domain is set in a two-dimensional $r-\theta$ plane polar mesh. The $\theta$ coordinate varies from 0 to $\pi$, and is split into 3200 zones. The radial coordinate is initiated with $R_{\rm{min}} = 0.006l$ and $R_{\rm{max}} = 0.061l$, and is split into 6400 zones. These resolutions capture both the radial and angular features well enough keeping the run time feasible. A smaller initial radial domain (as compared to the entire radial range of the problem) helps capture the shock at a higher Lorentz factor. The radial mesh eventually expands, moving the inner and outer boundaries using the moving mesh feature of \textit{the Jet Code}, and dynamically captures the entire radial range of the jets with temporal evolution. This is achieved by fixing the ratio of the inner to outer domain with respect to the shock front. We use a logarithmic time grid varying from 0.06 code units to 20 code units split in $10^5$ time steps.

We developed a code, \textit{Firefly}, to post-process the hydro simulations from the \textit{JET} code. The \textit{Firefly} code takes the complete two dimensional hydrodynamical evolution as its input and computes the three-dimensional synchrotron radiation using the standard afterglow model as described in the previous section. The 2D hydro data is extended to three dimensions by \textit{Firefly} using the rotational symmetry about the jet axis. The radial and $\theta$ resolution for the afterglow calculations assumes the same resolution as \textit{the Jet Code} output, while the $\phi$ axis is divided into 64 zones from 0 to $\pi$. The hydrodynamical parameters thus have the same value for all $\phi$ zones at a given ($r,\theta$). The hydro checkpoints are then binned over temporal resolution for the observer $t_{\rm{obs}}/dt_{\rm{obs}} = 0.03$. The binning ratio is a user choice, we choose the mentioned ratio to generate a smooth lightcurve without the loss of any physical features. \textit{Firefly} has three user input modes; to calculate the lightcurve, spectrum, or sky map. For each mode, it takes the total energy ($E_{\rm{tot}}$, interstellar nucleon density ($n_{\rm{\small{ISM}}}$), which is the same as $n_e$ in \ref {sec:TheoMod}), and other micro-physical parameters ($\epsilon_B, \epsilon_e$, spectral index $p$) as user inputs. The total energy for a two jet system is related to the isotropic equivalent energy by Eq. \ref{eq: Etot}. The quantities $E_{\rm{tot}}$ and $n_{\rm{ISM}}$ are scaled in \textit{the Firefly code} as their code unit values in the hydrodynamical evolution. Since in \textit{the Jet code} these values are scaled to unity in code units, the input values in \textit{the Firefly code} are the absolute values (corresponding to actual units, i.e. cgs or SI) pertaining to the problem. An observer is placed at luminosity distance ($d_L$) with a redshift ($z$) at an angle $\theta_{\rm{obs}}$. \textit{Firefly} can then be used to calculate the lightcurve (at any given observer frequency). \textit{Firefly} computes the cooling break and minimum break frequencies to correctly generate the broken power law spectrum of the GRB afterglow. It however does not account for the synchrotron self-absorption. 

\begin{equation}
    \label{eq: Etot}
        E_{\rm{tot}} = \frac{\theta_0^2}{2}E_{\rm{iso}}
\end{equation}

\begin{figure}
\centering
\includegraphics[width=0.3\textwidth]{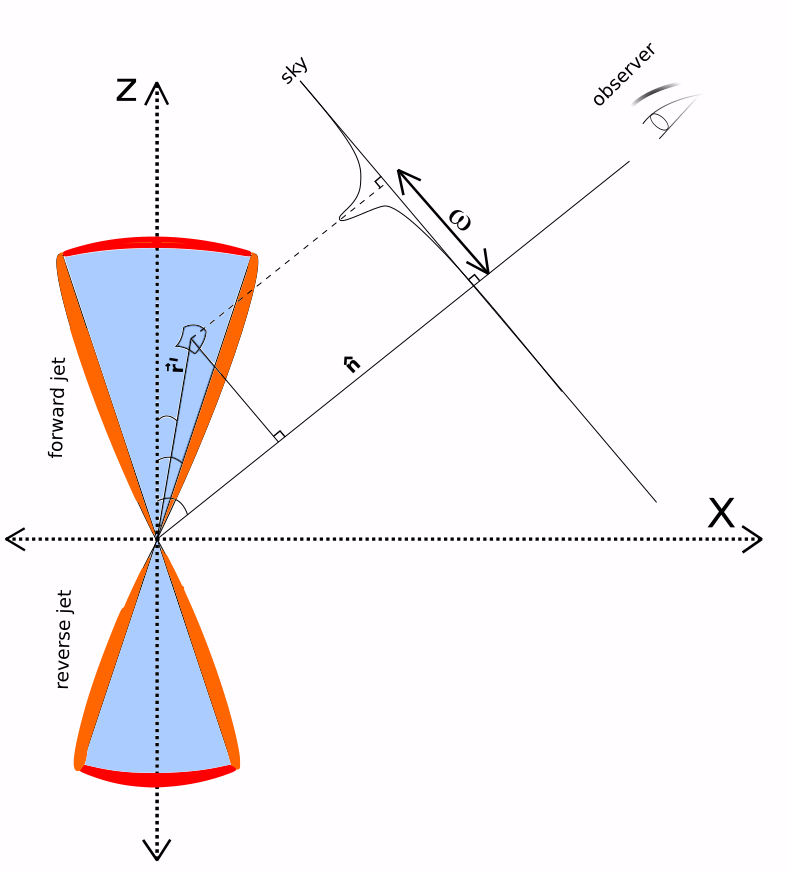}
\caption{\footnotesize{Pictorial representation of the sky map calculation. Contributions from a particular emission region ($\mathbf{\hat{r}}$) along the line of sight ($\mathbf{\hat{n}}$) are considered, and integrated over the entire domain. The line of sight is fixed at an angle of $\theta_{\rm{obs}}$. Its flux at the observer plane at $d_L \approx 40 Mpc$ from the merger is calculated.}}
\label{fig:sky_map_draw}
\end{figure}

Additionally, \textit{Firefly} can also be used to track the apparent motion of the object in the sky (as seen by the observer). This is achieved by computing the flux distribution (for a user-given frequency) along a plane perpendicular to the line of sight of the observer and the merger (hereon referred to as \textit{sky map}, see Fig. \ref{fig:sky_map_draw}), as a function of its distance from the merger center. This provides a more realistic approach to tracking the entire shock-ISM interaction region through the sky, and not just the jet head. This is especially relevant for off-axis observations and structured jets. The brightest emitting region from such a plot at multiple observer times can then be used to calculate the apparent motion of the object through the sky.

The sky map is computed by integrating the projected flux along a line on the plane perpendicular to the observer axis from different emitting regions of the jet-ISM interaction. To do this, we consider the x-z plane containing the jet axis and observer. A small emitting region of the jet at $\Vec{r}$ from the blast center is projected on this plane containing the jet axis and the vector ($\hat{n}$), connecting the blast center and the observer, by removing the $\hat{y}$ component (axis into the plane of the paper). This gives the position vector of the projected region:

\begin{equation}
    \Vec{r’} = \Vec{r} - (\Vec{r}\cdot\hat{y}) \hat{y}
\end{equation}

We then calculate its shortest distance ($\omega$) along the plane perpendicular to $\hat{n}$. This is given by the cross product of the remaining vector with the observer axis ($\hat{n}$).
$$\Vec{\omega} = \Vec{r’}\times \hat{n}$$. Since the resulting quantity is a vector, we project it along $\hat{y}$. The geometry can be expressed mathematically as Eq. \ref{skyloc}, and pictorially as Fig. \ref{fig:sky_map_draw}. This scheme is then repeated over the entire computation domain at a given observer time.

\begin{equation} \label{skyloc}
    \omega = \{[\Vec{r} - (\Vec{r}\cdot\hat{y})\hat{y}]\times\hat{n}\}\cdot\hat{y}
\end{equation}

\section{Results} \label{sec:Results}

\begin{figure}
\includegraphics[width=0.5\textwidth]{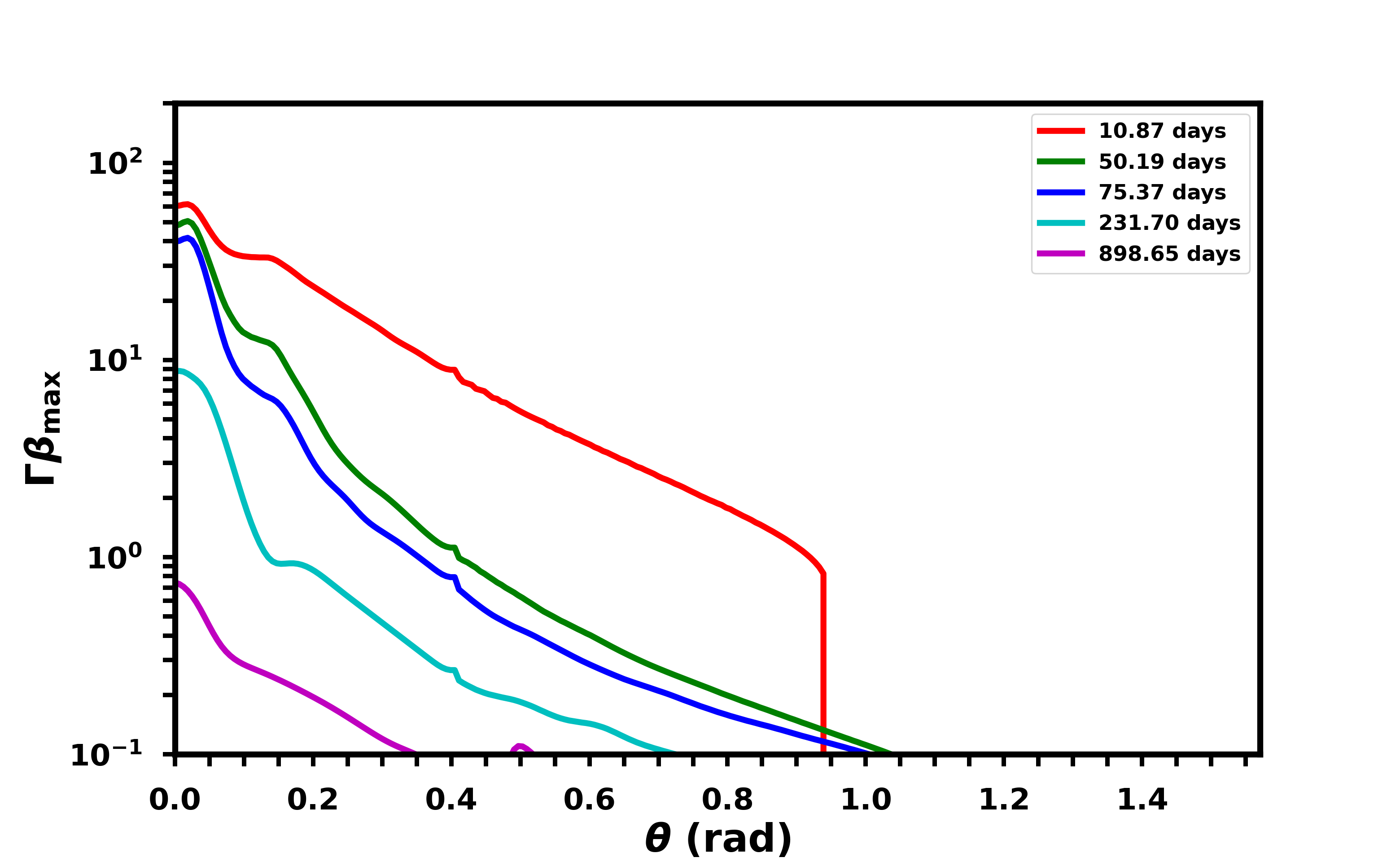}
\caption{\footnotesize{The four velocity evolution of the fireball. The red curve reflects the angular profile in the beginning of the simulation. The following colored plots are at various epochs under consideration in this paper. The boosted fireball starts with a narrow opening angle $\theta_{\rm{obs}} ~ 1/\Gamma_B ~ 0.25$ rad. It then spreads out as the fireball evolves.}}
\label{fig:angular_fourvelo}
\end{figure}

\begin{figure}
\includegraphics[width=0.5\textwidth]{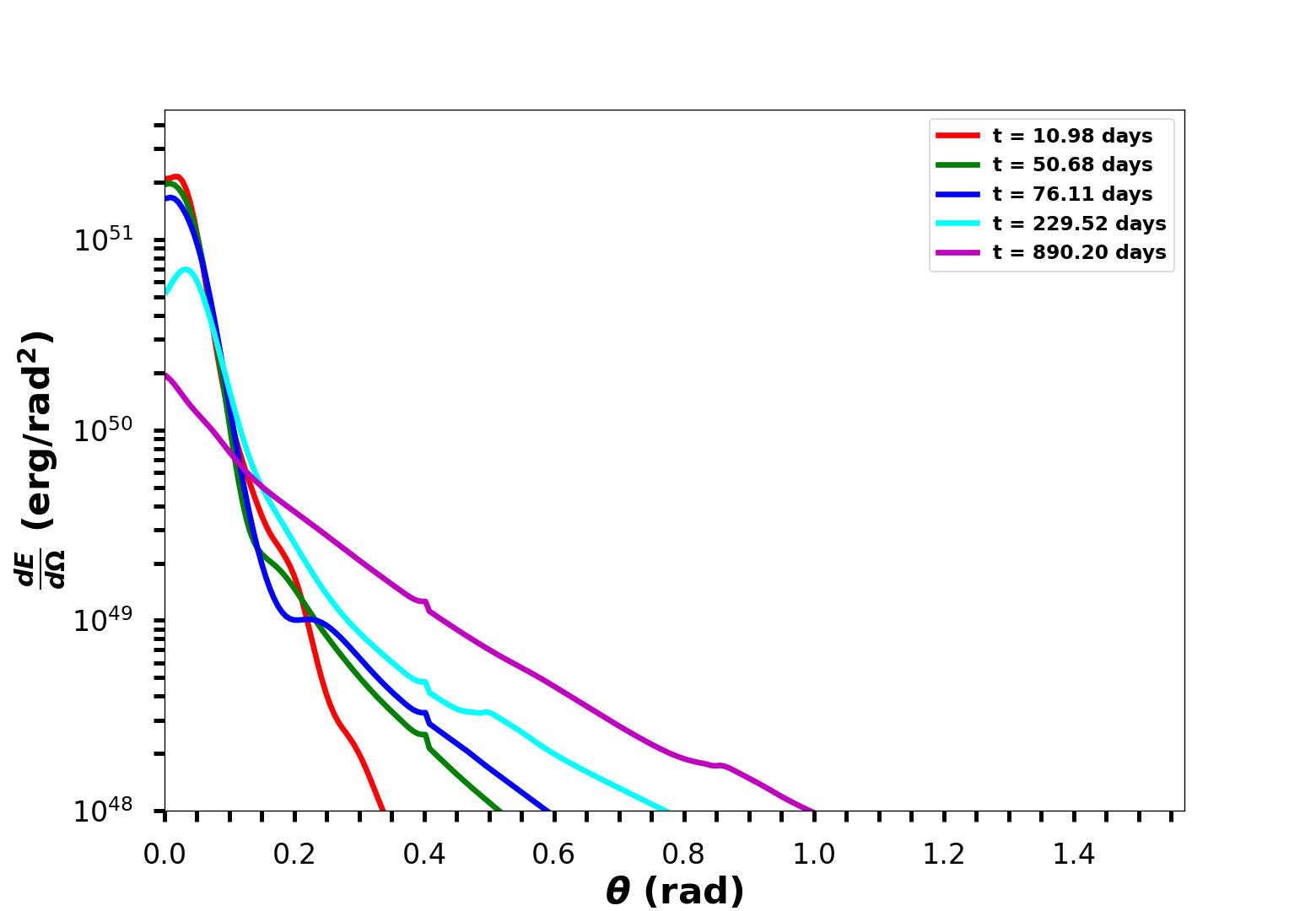}
\caption{\footnotesize{The same as Fig. 3. This figure shows the evolution of the boosted fireball energy profile. The energy per unit solid angle is plotted at various epochs.}}
\label{fig:jet_spread}
\end{figure}

 For this study, we explore the parameter space ${E_{iso},\gamma_B, \eta_0, \theta_{\rm{obs}}, n_{\rm{\small{ISM}}}, p, \epsilon_e, \epsilon_B}$. We carried out several runs of the hydrodynamical code for the boosted fireball model \citep{Duffell+2013} varying $\eta_0$, and $\gamma_B$. Values were selected from a parameter range of $\eta_0 \in [4,12]$ and $\gamma_B \in [4,12]$ (inspired by \cite{Wu+2018}). The temporal evolution of the angular energy profile and Lorentz factor for the fireball are plotted in Fig. \ref{fig:angular_fourvelo} and \ref{fig:jet_spread} respectively. We fixed the luminosity distance $d_L$ = 40 Mpc, which corresponds to a cosmological redshift of z = 0.00998. Since the Blandford-McKee solution for relativistic blastwave holds for most of the evolution, we fix the adiabatic index to 4/3.

\begin{figure}
\centering
\includegraphics[width=0.5\textwidth]{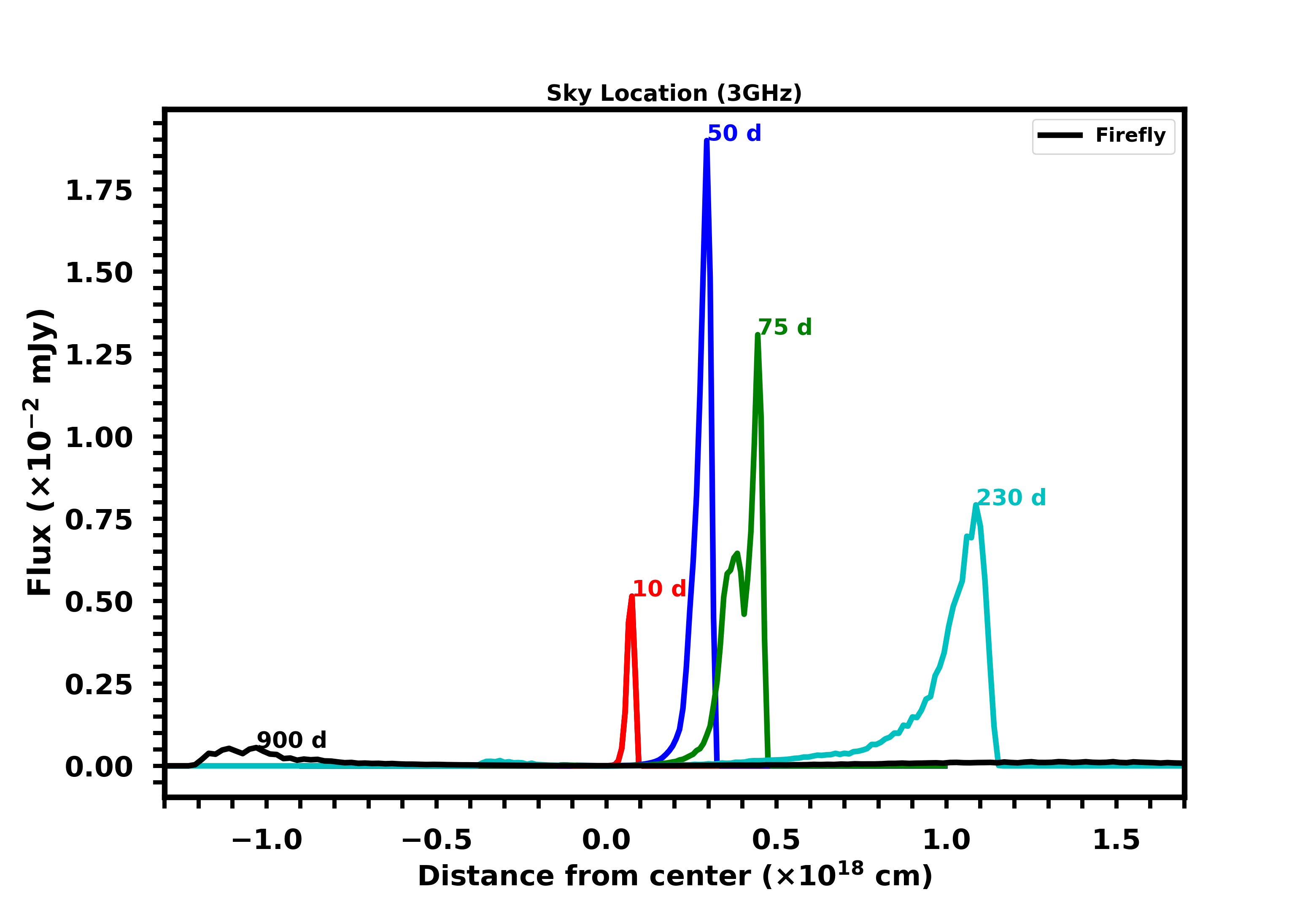}          
\caption{\footnotesize{Flux sky map for a boosted fireball model, generated using the \textit{Firefly} code. We have plotted the flux distribution along the plane perpendicular to the observer. The zero on the x-axis denotes the merger center. The flux map starts as sharp and near the merger. Around 900 days, the dominant region appears to come from behind the merger, a little over $10^{18}$ cm away from the center. This shows the emergence of the counter jet. The sky flux map is drawn for 5 instances in time, 10, 50, 75, 230, and 900 days.}}
\label{fig:sky_map_cj}
\end{figure}

 For a given set of hydrodynamics parameters, we fix all the microphysical afterglow parameters except $\theta_{\rm{obs}}$ and generate lightcurves as in Fig. \ref{fig:tp_tcj_multiangle}. We observe the late time bump in the lightcurve due to the counterjet as expected \citep{Li+2004,Granot+2003,Zhang+2013,Li+2019}. The counter jet excess flux of the afterglow lightcurves indicates a similar re-brightening timescale for off-axis observations, except for far off-axis ($\gtrsim 1.2$ rad). For these large off-axis observer angles, the re-brightening initiates at earlier times. Fig. \ref{fig:tp_tcj_multiangle} shows the same. This also shows that the second bump is indeed associated with the counter jet. Emission from the counter jet shares the same profile in the afterglow lightcurve as the forward jet, but with a very significantly delayed $t_{\rm{obs}}$, due to its orientation pointing and moving away from the observer. While all of the initial afterglow is due to the forward jet, once the jet spreads out and is spherical enough, the counter-jet eventually becomes visible to the observer. Flux from the counter jet gradually increases as the forward jet did, and it eventually outshines the forward jet. This occurs because the observer time $t_{\rm obs}$ for the counter-jet corresponds to an earlier lab-frame time than for the forward jet.  The forward jet eventually spreads out and slows down, leading to a steeply declining lightcurve, while the counter jet is still effectively beamed and relativistic at the same observer time. For a brief period, this leads to a higher flux from the counter jet. This effect is also captured in the sky map $\sim 900$ days, Fig. \ref{fig:sky_map_cj}. The flux from the counter jet also eventually peaks and is briefly brighter than the forward jet before it declines. The flux from both the jets converge asymptotically and eventually contribute the same flux as the jets spread and eventually become an isotropic system.

\subsection{Counter-jet time}\label{subsec:Countetjettime}
While all of the initial afterglow is due to the forward jet, once the jet spreads out and is spherical enough, the counter-jet eventually becomes visible to the observer. This can be seen as a late-time bump in the afterglow lightcurves \citep{Li+2004,Granot+2003,Zhang+2013,Li+2019}, and Fig. \ref{fig:lightcurve}. \cite{Hajela+2022,Li+2019} gives an estimate of the counter-jet visibility time as

\begin{equation}
    t_{cj} \approx (1+z)t_{NR} \approx 1900(1+z)(E_{iso,53}/n_0)^{1/3} \text{days}
\end{equation}

\begin{figure}
\includegraphics[width=0.5\textwidth]{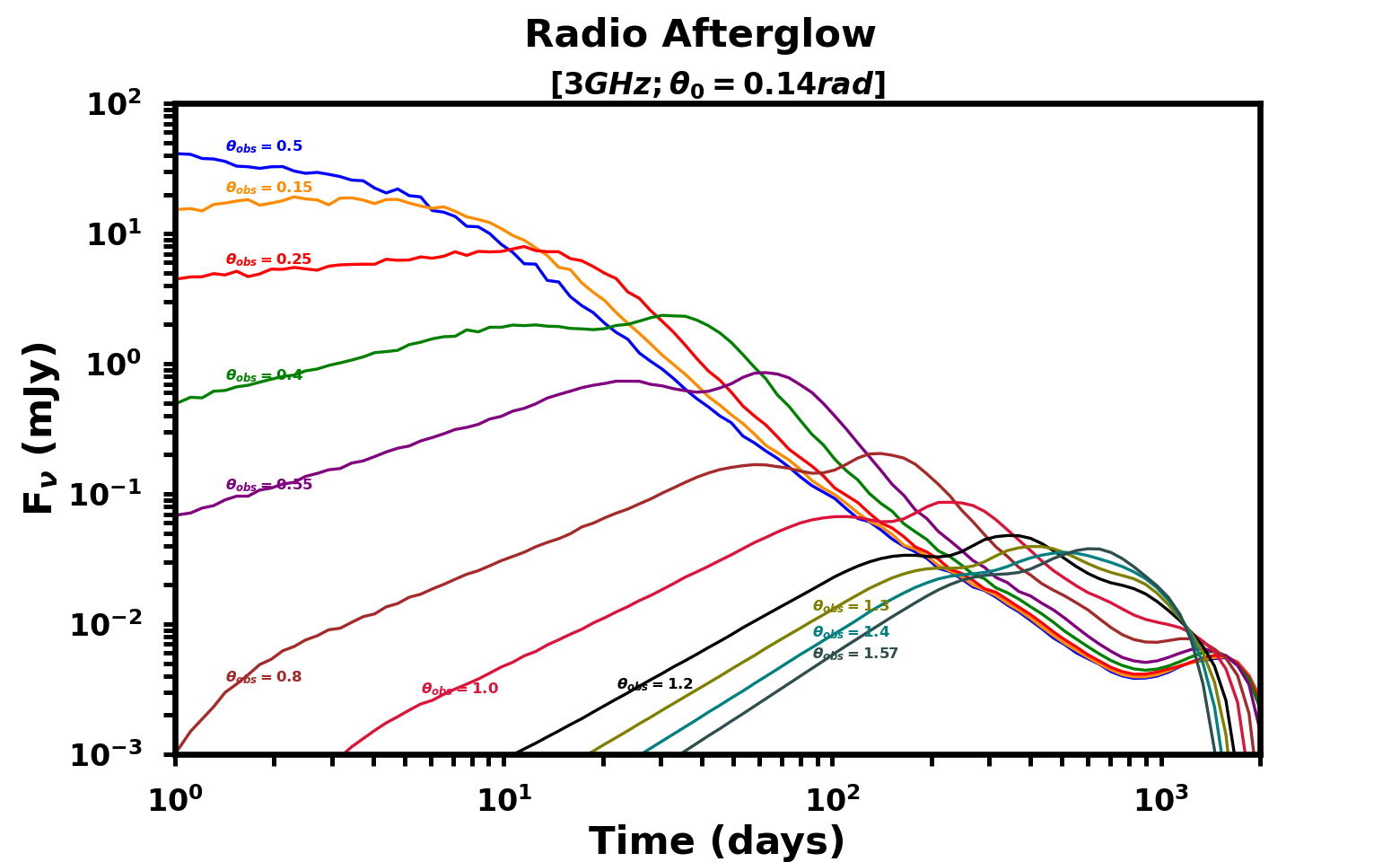}
\caption{\footnotesize{Lightcurves for a given jet-ISM parameters, but varying observer angle. The lightcurves for observer angles, $\theta_{\rm{obs}} = [0.5,0.15,0.25,0.40,0.55,0.80,1.00,1.20,1.30,1.40,1.57]$ rad are shown. For $\theta_{\rm{obs}} \textless \theta_0 (=0.14 rad)$, we see the lightcurves start with a plateau like feature. For larger viewing angles ($\theta_{\rm{obs}} \textgreater \theta_0$), The lightcurve increases first, due to emissions from the jet limbs. As the jet slows down and spreads, the jet core eventually becomes visible, and lightcurve peaks. Irrespective of the observer angle, the second rise and peak, due to the counter jet happen almost at the same time around 1000 days. For extreme off-axis ($\theta_{\rm{obs}} \textgreater 1.2$ rad), we enter a regime when the forward and counter jets} can not be distinguished from each other. This results in single peak light curves. These lightcurves peak at a much later time, closer to the second re-brightening, and fall off sharply almost immediately.}
\label{fig:tp_tcj_multiangle}
\end{figure}

It must be noted here that since we are off-axis from the jet core, we do not need the entire jet and counter-jet to be completely spherical for the counter-jet to become visible. We model the time scale for the counter-jet peak emission ($t_p^{cj}$), in terms of the time scale of the forward jet peak emission($t_p^j$), and the observer angle ($\theta_{\rm{obs}}$).

\begin{figure}
\includegraphics[width=0.5\textwidth]{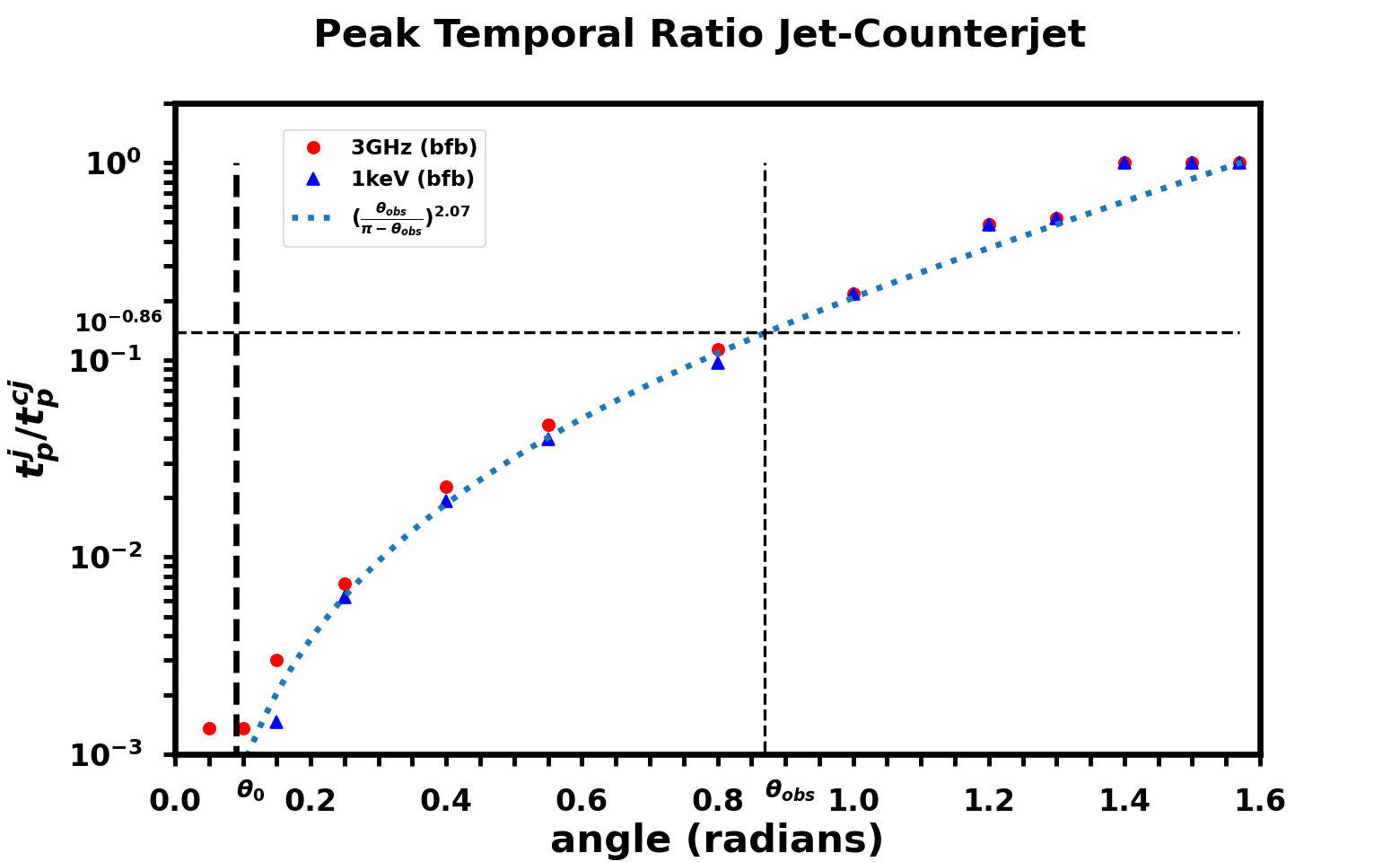}
\caption{\footnotesize{Ratios of the time scales when the emissions from the forward jet peak, to that of the counter jet on the y-axis. Compared to the observer angle along the X-axis. This is measured at two frequencies, 3GHz radio band, and 1keV X-ray band. We find no difference in their dependence on the observer angle, as expected. This temporal ratio follows an empirical scaling law Eq.\ref{TemporalPeakRatio}. This is plotted as blue dashed lines. The region of interest when $t_p^{j}/t_p^{cj}\sim 175/1234$, which is the prospective observer viewing angle $\theta_{\rm{obs}} \sim 0.85$ rad,}, is also marked with thin black dashed lines. The thick dashed line at $\theta_0 = 0.09$ shows the jet opening angle. Due to various shocks and numerical instabilities at the jet wing boundaries, the temporal ratio is slightly different for X-ray and radio near the jet opening angle. Since $\theta_{\rm{obs}}$ is much greater than the jet opening angle, this numerical effect can be ignored without loss of generality.}
\label{fig:temp_pr}
\end{figure}

For a given observer angle at a given frequency, we identify the time and flux from the two peaks due to the forward and counter jet from the lightcurve. We then carried out a parametric study for the temporal ratio of these two epochs over a range of $\theta_{\rm{obs}} \in [0,\frac{\pi}{2}]$. The results are shown in Fig \ref{fig:temp_pr}. The same scheme is used for two different frequencies, 3GHz, and 1keV lightcurves to study the chromatic dependence as well.

We find that the temporal peak ratios have achromatic behavior. That is the time scale relations between the jet and counter-jet peak emission should be valid for all frequencies. For an on-axis observer, within the jet opening angle $\theta_0$ (\textgreater $\theta_{\rm{obs}}$), the forward jet peak time is the earliest observer time, hence the ratio $t_p^j/t_p^{cj} \ll 1 $. While on the other extreme for far off-axis observer, $\theta_{\rm{obs}} \gg \theta_0$, the forward and counter jet are indistinguishable, and the ratio $t_p^j/t_p^{cj} \approx 1$. That is the total emission is equally contributed by both the jets, leading up to a single peak. In between, for an off-axis observer $\theta_{\rm{obs}} > \theta_0$ ($\theta_{\rm{obs}} \approx \pi/2$), the peak flux from forward jet gradually moves towards the peak from the counter jet as we go further off-axis. This is shown in Fig. \ref{fig:tp_tcj_multiangle}. In this regime, assuming $\gamma_B$, $\eta_0$ and $n_{\rm{ISM}}$ remains fixed, we can fit a curve relating the peak time ratios to the observer angle as:

\begin{equation} \label{TemporalPeakRatio}
    \frac{t_p^j}{t_p^{cj}} = \left[\frac{\theta_{\rm{obs}}}{\pi - \theta_{\rm{obs}}}\right]^{2.07}
\end{equation}

With $t_p^j = 175 d$, and assuming the counter-jet peak emission is later than 1234 d, we can place a lower bound of $\frac{t_p^j}{t_p^{cj}} > 10^{-0.85}$ from observations.  Using our scaling law Eq. \ref{TemporalPeakRatio}, this gives $\theta_{\rm{obs}} \approx 49.8^\circ$. 

The ratio of fluxes at $t^j_p$ and $t^{cj}_p$ do not follow such a simple relationship. Since the flux is also dependent on the break frequencies, the flux ratio at the jet counter-jet peak epochs need not be achromatic. Rather they should depend strongly on the spectral breaks. In the particular case of GRB170817A, the spectral index maintains a rather fixed value, they happen to align.

\subsection{Lightcurve} \label{sec:Lightcurve}
Since GRB170817A was observed off-axis, the lightcurve rises and peaks around 160 days. This is because early emissions arise from the off-axis jet material, and higher-energy material continues to come into view as the jet decelerates.  The lightcurve peaks when the jet core becomes visible to the observer. The flux starts to decline thereafter. This is due to the fact the jet slows down and loses energy as it
expands. Up until this point, the radiation is dominated by the forward jet oriented towards us. Hydro simulations with a single jet or double jet can not be
distinguished from their afterglows until much later.

The lightcurves are constructed using the Standard Afterglow theory detailed in previous section. We use our code \textit{Firefly} for this. We construct various lightcurves and constrain the afterglow microphysical parameters with the observations at $\theta_{\rm{obs}} \approx 49.8^\circ$. We find the best match for the parameters reported in Table \ref{t:parameters}. Taking into account the recent observations ($t_{\rm{obs}} \textgreater 900$ days) we find that the excess flux observed could be associated with the counter jet re-brightening for GRB170817A, if one only considers the lightcurve.

\setlength{\tabcolsep}{20pt}
\renewcommand{\arraystretch}{1}
\begin{deluxetable}{|c|c|}
\tablewidth{0pt}  
\tablecaption{Parameter Range for Fireball Jet} 
\tablehead{ \colhead{Parameter} & \colhead{Double Fireball} }
\startdata
\hline
\hline
$\gamma_B$ & 8.0\\
\hline
$\eta_0$ & 6.5\\
\hline
$\theta_0$ & $7.2^\circ$\\
\hline
$E_{tot}$ & $1.6\times 10^{49} erg$\\
\hline
$E_{iso}$ & $1.98\times 10^{51} erg$\\
\hline
$\theta_{obs}$ & $49.8^\circ$ \\
\hline
$n_{ISM}$ & $0.1 cm^{-3}$\\
\hline
p & $2.13$\\
\hline
$\epsilon_e$ & $0.4$\\
\hline
$\epsilon_B$ & $10^{-3}$\\
\enddata
\label{t:parameters}
\tablecomments{\footnotesize{There are eight free parameters in the boosted fireball afterglow model. The first two for the jet itself for boosted fireball model ($\gamma_B$, $\eta_0$). The remaining six parameters come from the standard afterglow model. The parameter range for them is chosen according to the physical limits set on them. These are the values for which we find the best agreement between the 3GHz and 1keV lightcurve. These are the parameters required to correlate the 1keV excess flux to the counter jet emission for GRB170817. Most notable difference from the previous studies are the values of $n_{\rm{ISM}}$, and $\theta_{\rm{obs}}$.}}
\end{deluxetable}

\begin{figure*}
\centering
\includegraphics[width=0.6\textwidth]{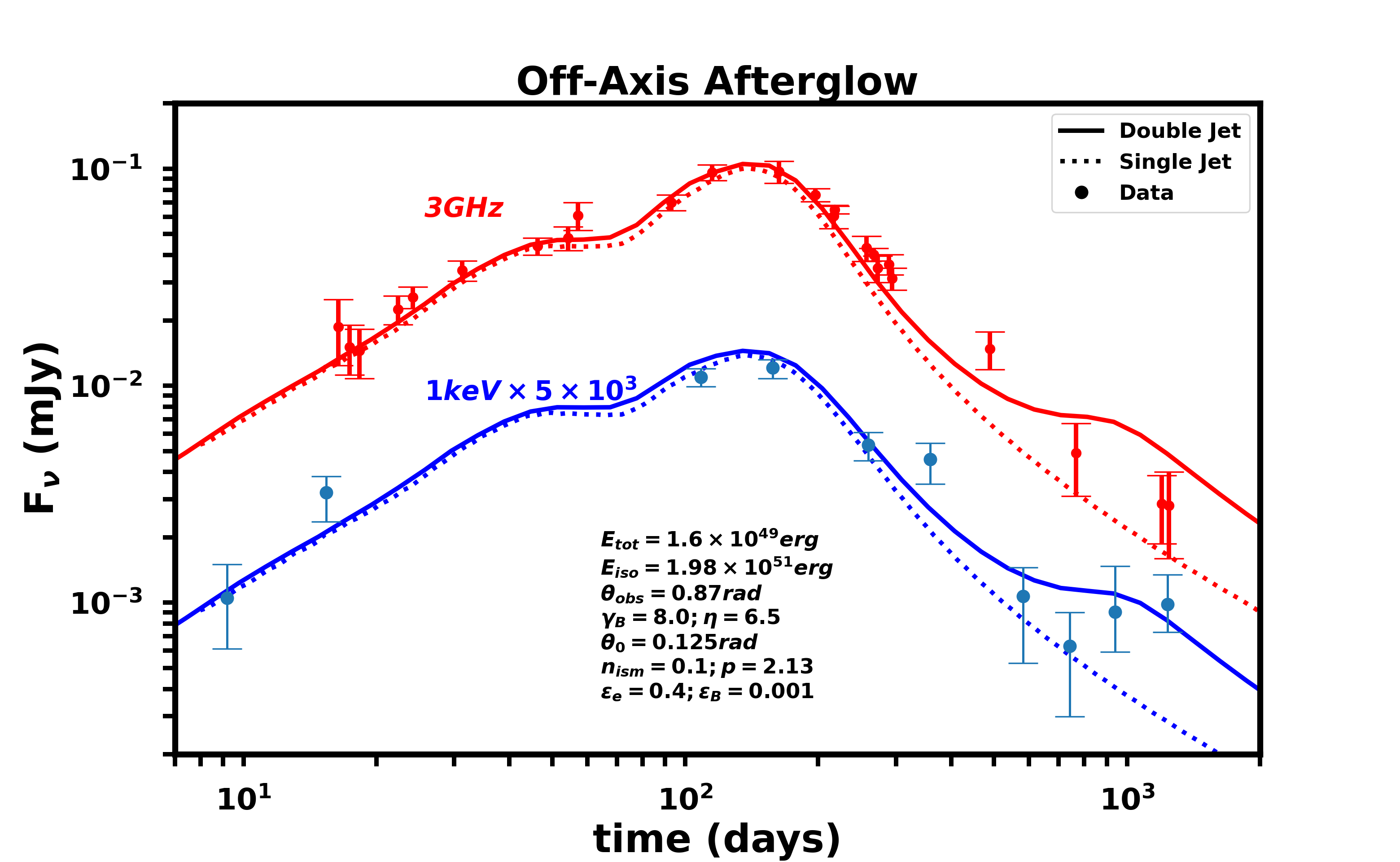}
\caption{\footnotesize{Lightcurves at 3GHz (red) and 1 keV (blue) for GRB170817A as observed from $\theta_{\rm{obs}} = 49.8^\circ$. The solid circles are data taken from various sources listed in the text. The solid and dashed lines are lightcurves with and without the counter jet respectively. The clear late-time excess seen in the solid lines, compared to the dashed line, implies the excess emissions come from the counter jet, peaking around 1000 days. The excess flux seen in the data, especially for the 1keV X-ray band matches very closely with the counter jet re-brightening within the errors of observation. But a stronger emission is expected at 3 GHz around the re-brightening time.}}
\label{fig:lightcurve}
\end{figure*}

We focus our results mainly at two frequencies, X-Ray on 1keV and 3GHz radio band. The frequencies are chosen to compare our results with the observations reported in \citep{Hajela+2022, Troja+2021}. Fig. \ref{fig:lightcurve} shows the comparison. \cite{Hajela+2022} and the models discussed therein, use the afterglow observations till 700 days to constrain the jet and its afterglow parameters.

Fig. \ref{fig:lightcurve} shows that it is possible to explain the excess flux observed for GRB170817A, as emissions originating from the counter-jet. It also shows how the lightcurve would look if only a single forward jet was considered. The main deviation from the previously accepted set of parameter values for the sGRB jet is in the observer angle. We find a far off-axis observer is needed to explain the early re-brightening from the counter jet. We find an observer located at $49.85^\circ$, along with the set of parameters reported in table \ref{t:parameters}, would observe the emissions from the counter jet at around 1234 days at 1keV, coincident with the excess flux. Along with the same lightcurve for previous times.

While the agreement for 1keV lightcurve is convincing, we see around the same time 3GHz does not show a significant excess in flux. As we previously argued, we expect the counter jet re-brightening of the system should be seen across all frequencies. One explanation can be that by this time, the synchrotron self absorption breaks lie beyond 3GHz. It is possible that the emissions below self-absorption break from the counter jet might also undergo more loss, diminishing its contribution as compared to the forward jet. This is due to the fact the emissions experience more ISM interaction as it passes through both the counter jet and jet lobes. Unlike the emissions below self-absorption from the forward jet, which get self-absorbed only in the forward jet lobe. We find this break using Eq. \ref{eq:sync_self_absorb} to be $3.6\times 10^8$ Hz around $t_d \sim 1000$ days. This does not explain the lower flux observed at 3GHz, compared to predictions by the counter jet (Fig. \ref{fig:lightcurve}). We note that this discrepancy remains. However, \cite{Troja+2021} found no such excess flux at 5keV band as well.

\subsection{Spectrum} \label{sec:Spectrum}
Along with the afterglow, the spectrum of GRB170817A has been closely observed.  The power-law index $p$ characterizing this spectrum is expected to evolve as the blastwave transitions from a highly relativistic to non-relativistic regime \citep{Bell+1978, Blandford+1978}.  However, no such major variation has been seen \citep{Hajela+2022}. Thus, all models so far have assumed a fixed-\textit{p} value \citep{Wu+2018, Troja+2021, Hajela+2022}, and references therein). Re-evaluation of values of \textit{p} \textgreater 2.166, i.e. larger than the best fit value for t \textless 900 days can be ruled out \citep{Hajela+2022}.

For our study, we assume a non-evolving value of $p$. We fix the parameters from the afterglow and we generate the corresponding spectrum using our code \textit{firefly}. \textit{Firefly} however does not include synchrotron self-absorption. 

We find the best fit at $p = 2.13$. This is similar to other results from the boosted fireball model, p = 2.154 reported by \cite{Wu+2018}. Which is approximately the same as $p = 2.166 \pm 0.026$ from the latest epochs \citep{Hajela+2022, Troja+2021}. Another notable feature of the spectrum is the same power law behavior over the entire spectrum from radio to X-ray. That is none of the frequency breaks lie within the range $3\times 10^6$ Hz to $3\times 10^{17}$. While the synchrotron self absorption is not captured in our code, we estimated the value to be around $3.6\times 10^8$ Hz following Eq. \ref{eq:sync_self_absorb} \citep{Granot+2002}. Which lies below the frequency range of our interest.

\begin{equation}
\label{eq:sync_self_absorb}
    \nu_a = 3.59(4.03-p)e^{2.34p}\left(\frac{\epsilon_e^{4(1-p)}\epsilon_B^{p+2}n_0^4E_{52}^{p+2}}{(1+z)^{(6-p)}t_d^{3p+2}}\right)^{\frac{1}{2(p+4)}} Hz
\end{equation}

It must be noted at this point that our parameters were chosen such that we can coincide the excess flux observed for GRB170817A, with the rise of counter-jet emission. While this may seem an ad hoc way of parameter estimation, our parameters are well within the permissible limits of the microphysics behind the afterglow theory.

\subsection{Apparent Motion} \label{sec:AppMotion}
\begin{figure}[htb!]
\includegraphics[width=0.5\textwidth]{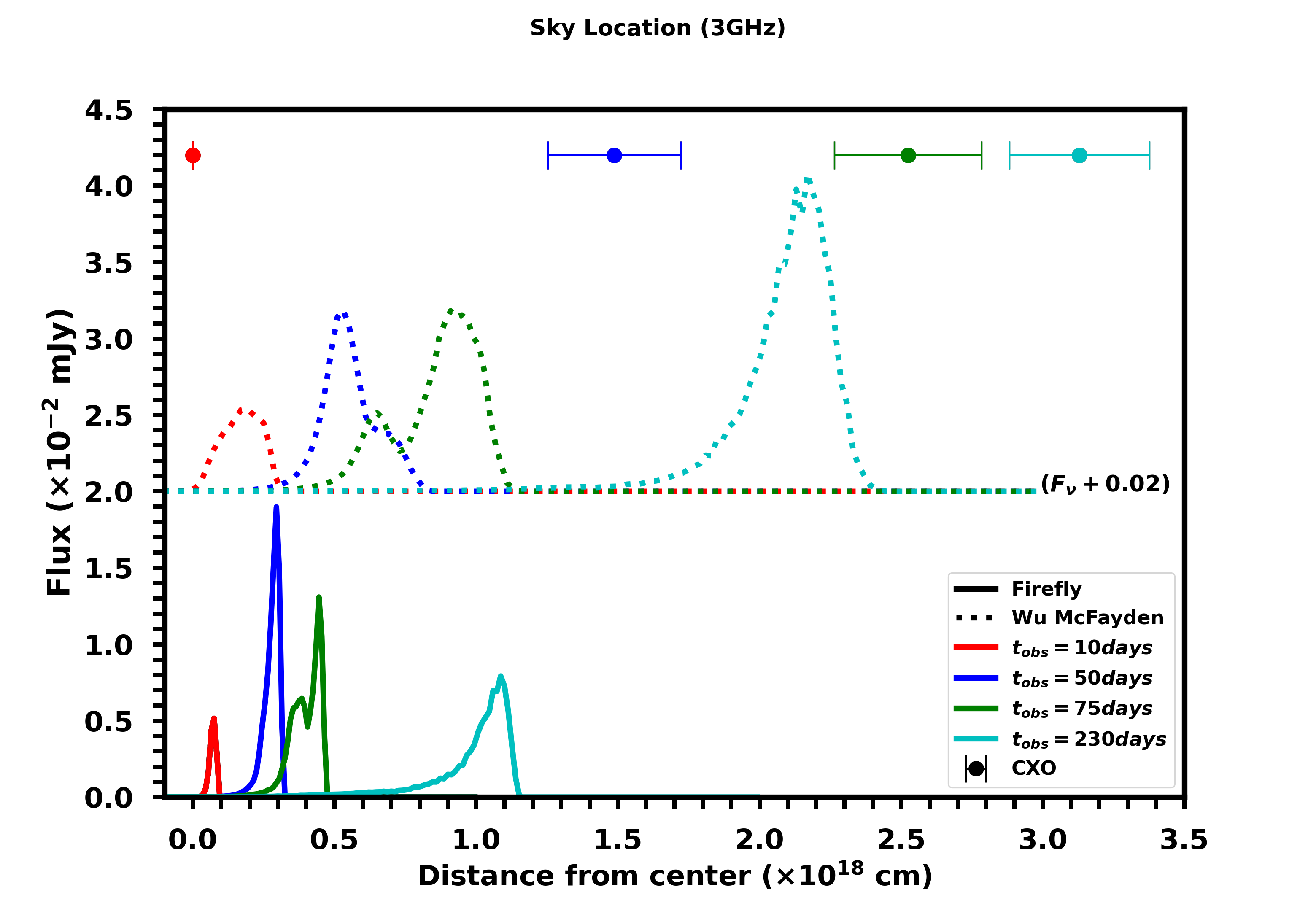}
\caption{\footnotesize{The location of the maximum emitting region in 3 GHz for the off-axis jet of GRB170817A is plotted. The observer is placed at $\theta_{\rm{obs}} = 0.87$ rad. Plotted at the top (the solid circles), are the observed physical distance of the merger center, to the maximum emitting region on the sky in 3GHz, from \cite{Mooley+2018}. Plotted below that is the flux map at the different epochs. These show the apparent motion of the flux contributions from the afterglow, along the line joining the merger center and the observer. The solid line are using the parameters from this paper (\ref{t:parameters}). While the dashes line are that using the \cite{Wu+2018} model. The peaks from these plots are used to identify the location of GRB170817A for that give time.}}
\label{fig:sky_loc}
\end{figure}

Using the sky map feature of our \textit{Firefly} code, we plot the flux versus sky location. Fig \ref{fig:sky_loc} shows this output at various observer times when observed at 3 GHz. For a given time (say blue curve in Fig \ref{fig:sky_loc}), we see a narrow profile with a sharp peak. The peak represents the brightest spot for GRB170817A afterglow in the sky, and hence its observed distance from the merger center. Over time the profile spreads out and moves away from the merger center. The flux at the peak also decreases with time and follows the lightcurve. The spread in the profile is directly correlated with the increase in shock width as the jet spreads and decelerates ($\Delta R \sim 1/\Gamma^2$).

Eventually, at a later time $\sim 900$ days, we see the maximum flux in the opposite direction, as if the jet switches its location. This is from the now brighter counter jet, which is still beamed and relativistic. The exact time at which this happens depends on the observer angle as discussed previously. This is in agreement with our previous argument that the late time excess flux is primarily due to the counter jet.

For now, we consider the epochs before such an excess flux. Taking the peaks as the jet location corresponding to that time, we can calculate the apparent motion from Fig. \ref{fig:sky_loc} directly. Table \ref{t:slyloc} summarises the results for apparent motion. We find the apparent velocity remains superluminal even at 230 days. However, the apparent velocity decreases from ~2.1c to 1.6c through 200 days since the BNS merger.

\setlength{\tabcolsep}{2pt}
\renewcommand{\arraystretch}{1.5}
\begin{deluxetable}{|c | c | c|}
\tablewidth{0pt}  
\tablecaption{Apparent motion for GRB170817A with Firefly} 
\tablehead{ \colhead{ Time since merge} & \colhead{Distance from merger center} & \colhead{Apparent velocity}\\
\colhead{(days)} & \colhead{$(\times 10^{16} cm)$} & \colhead{$(cm s^{-1})$}}
\startdata
\hline
\hline
10 & 7.5 & \\
\hline
50 & 29.5 & 2.122 c\\
\hline
75 & 44.5 & 2.354 c\\
\hline
230 & 108.7 & 1.6 c\\
\enddata
\label{t:slyloc}
\tablecomments{\footnotesize{This table summarizes the apparent motion for our boosted fireball model jet, following the parameters mentioned in Table \ref{t:parameters}. The third column is the calculated apparent motion, in terms of speed of light 'c', calculated between the two epochs mentioned in the first column.}}
\end{deluxetable}

Our inferred superluminal motion conflicts with observations. The apparent motion is highly sensitive to the observer angle \citep{Ryan+2023}  and \cite{Mooley+2018} observed a superluminal apparent velocity of seven times the speed of light for the GRB170817A, even after a year. Centriod corrected fits for apparent motion give $\theta_{\rm{obs}} \sim 20^{\circ}$ for jet opening angle $\sim 3.5^{\circ}$ \citep{Ryan+2023}. This hints at a smaller observer angle than what our model predicts. Since the superluminal motion observations are a direct result of the orientation of the jet with respect to the observer, we put a stronger emphasis on the observed motion and its consequences. 

\cite{Wu+2018} did not compute the apparent motion of the jet with time for their models.  However, we re-ran their setup using the parameter values from their best-fit model, to determine the superluminal motion for the \cite{Wu+2018} setup.  This is also shown in Figure \ref{fig:sky_loc}.  For the setup of \cite{Wu+2018}, we see an improved match for the apparent velocity with a smaller observer angle. However, Fig. \ref{fig:sky_loc} suggests that their setup is also inconsistent with the data, suggesting that an even smaller viewing angle than $26.9^\circ$ may be necessary to match all available data.

\section{Conclusion}  \label{sec:conclusion}
GRB170817A is one of the most crucial transients in recent years. While it was the first off-axis sGRB afterglow observed it has reignited multiple questions in the field of gamma-ray bursts and afterglow. The afterglow from GRB170817A has also shown a significant agreement with a structured jet interaction with a constant ISM. In this study, we re-evaluated the parameters to investigate the conditions under which the excess flux could be due to the counter jet.

We ran multiple hydrodynamic simulations for a double sided boosted fireball jet model, varying $\gamma_B$, and $\eta_0$. The first peak in the afterglow lightcurves was used to constrain the jet opening angle at $\theta_0 = 1/\gamma_B = 0.125$ rad ($\gamma_B = 8.0$). The fireball spread parameter, quantified by $\eta_0$ was similarly constrained to be 6.5. \cite{Wu+2018} find asymptotic Lorentz factor $\gamma_B \sim 11$ and $\eta_0 \sim 8$ (Fig. \ref{fig:sky_loc}). Fixing $\gamma_B$, and $\eta_0$, we constructed lightcurves at 3 GHz radio and 1 keV X-ray for 11 observer angles each. In our model, there is no additional source for excess flux. Thus the second peak observed in Fig. \ref{fig:lightcurve} is purely due to the counter-jet interactions. We also observe that at later times, since the counter jet has longer $t_{\rm{obs}}$, it becomes brighter than the forward jet. After this turnover time, the contributions from the counter jet leads to the second peak in the lightcurves. We then computed an empirical scaling law between the jet and counter-jet peak emission time scales, with the observer angle Eq. \ref{TemporalPeakRatio}.

For the excess flux of GRB170817A to correspond with the emissions from the counter-jet, we can place a lower bound of $\frac{t_p^j}{t_p^{cj}} > 10^{-0.85}$ from observations (sec \ref{subsec:Countetjettime}). Comparing simulations with the observations (Fig. \ref{fig:lightcurve}) we constrain our observer angle to be $49.8^\circ$. In contrast, previous attempts studies have found $\theta_{\rm{obs}} \sim 27^\circ$ \citep{Resmi+2018, Lazzati+2018, Mooley+2018a, Wu+2018}. Using the relation Eq. \ref{TemporalPeakRatio} we estimate the peak of this second component to occur around 1268 days since explosion. The afterglow lightcurve shall then decay sharply, with a temporal slope faster than before the excess started.

Using this observer angle, we fit the 3 GHz and 1 keV lightcurves, to constrain the microphysical parameters associated with the standard afterglow model. The best match between observations and our calculations from the Firefly code fixed the parameters $\epsilon_e = 0.4$, and $\epsilon_B = 0.001$. \cite{Resmi+2018} and \cite{Wu+2018} report $\epsilon_e \sim 0.2$ and 0.3 respectively. Further typical values for $\epsilon_B$ lie within the range $\sim 10^{-5} - 10^{-1}$ (\cite{Wu+2018} and references therein). Thus, we find both afterglow parameters agree closely with previous attempts at modelling this afterglow. While the inherent degeneracy of $E_{iso}/n_{\rm{\small{ISM}}}$ remains, owing to the Blandford-McKee solution of blastwave. To break this degeneracy, we fix our isotropic equivalent energy closer to the realistic scales of such BNS mergers. We fix $E_{iso} \approx 1.98\times 10^{51}$, with total jet energy $E_{tot} = 1.6\times 10^{49}$. This constraints the circumburst medium density $n_{\rm{\small{ISM}}} = 0.1$/cm$^3$. The value is significantly much larger than $ 10^{-5} - 10^{-3}$, suggested by previous studies \citep{Margutti+2018,Mooley+2018a, Lazzati+2018}, but in agreement with the expected value for such a medium. In contrast, \cite{Wu+2018} fix $ n_{\rm{\small{ISM}}} = 10^{-3}$ and obtain $E_{\rm{tot}} = 2\times 10^{49}$ erg for a similar model. Table \ref{t:parameters} summarizes the parameter values obtained.

Narrowing down all the parameters for the jet-ISM interaction, we then used the sky map feature of Firefly. Tracking the most luminous region of 170817A afterglow along the plane perpendicular to the observer. We get an apparent velocity of the afterglow through the sky as 2.12c in the initial days, slowing down to 1.6c later after the first peak. However, \cite{Mooley+2018} report the apparent velocity observed as seven times the speed of light. This contradicts our model and implies a smaller viewing angle. \citep{Wang+2024} also found larger viewing angle ($\sim 50^{\circ}$) fit to the afterglow, which reduced to $\sim 18.16^{\circ}$ after correcting for the superluminal motion. Other attempts at modeling the re-brightening with $\theta_{\rm{obs}} \sim 27^\circ$, include power law momentum distribution in the kilonova \citep{Kathirgamaraju+2017}, a fast moving tail in the dynamical ejecta and central engine powered radiation from the compact object (see, \cite{Hajela+2019} and references therein for detailed discussion).

We conclude that it is possible to associate a re-brightening time with the counter jet visibility time and constrain the observer angle from that. For GRB170817A/ GW170817 we can further fine-tune the model to match the radio and X-ray lightcurves for that observer angle.  However, although the association of counter jet visibility with excess flux time scale and lightcurve matching indicates a possible solution to the X-ray excess, and not for the radio band. Only by comparing the apparent motion of the object through the sky, we can nullify this seeming correlation. Hence we propose,that jet and counter jet re-brightening timescales can also be used to constrain the observer's orientation with respect to the jet. Further, full diagnostics including the apparent (superluminal) motion along with parameter estimation with lightcurve fitting is required to narrow down the geometrical orientation for such beamed emissions observed off-axis.

\begin{acknowledgments}

We acknowledge D. Giannios, H. Wang, G. Ryan and H. van Eerten for their helpful comments, and H. van Eerten for also helping set up Boxfit. We also acknowledge A. Hajela for discussions on the new data from GW170817A.  Hydrodynamical calculations were carried out on the Petunia cluster at Purdue University.  This project and the development of the code \textit{Firefly} was supported by NASA under grant No. 80NSSC22K1615.
\end{acknowledgments}

\typeout{}
\bibliography{rana}

\end{document}